\documentclass[prd,twocolumn,floatfix,nofootinbib,superscriptaddress,showpacs]{revtex4}
\pdfoutput=1
\usepackage{graphicx}
\usepackage{epsfig}
\usepackage{bm}
\usepackage{amssymb}
\usepackage{float}
\usepackage{amsmath}
\usepackage{dcolumn}
\usepackage{tensor}

\usepackage{dcolumn}
\usepackage{cancel}
\usepackage[colorlinks]{hyperref}
\usepackage[usenames,dvipsnames]{color}
\hypersetup{
     breaklinks=true,
    pdfstartview={FitH},    
    colorlinks=true,       
    linkcolor=blue,          
    citecolor=red,        
    filecolor=magenta,      
    urlcolor=blue,           
    anchorcolor=green,      
    linktocpage=true
}

\usepackage{amsmath,latexsym}
\usepackage{placeins}
\usepackage[toc,page]{appendix}

\newcommand{\mathsym}[1]{{}}
\newcommand{\unicode}[1]{{}}
\newcommand{\be}{\begin{equation}}
\newcommand{\ee}{\end{equation}}
\newcommand{\bea}{\begin{eqnarray}}
\newcommand{\eea}{\end{eqnarray}}
\newcommand{\beaa}{\begin{eqnarray*}}
\newcommand{\eeaa}{\end{eqnarray*}}

\begin{document}

\title{Big Bang Nucleosynthesis constraints on $f(T,T_G)$ gravity}

\author{Petros Asimakis}
\affiliation{Department of Physics, School of Applied Mathematical and Physical Sciences, National Technical University of Athens, 9 Iroon Polytechniou Str., Zografou Campus GR 157 80, Athens, Greece}

\author{Emmanuel N. Saridakis}
\affiliation{National Observatory of Athens, Lofos Nymfon, 11852 Athens,
Greece}
\affiliation{CAS Key Laboratory for Researches in Galaxies and Cosmology,
Department of Astronomy, University of Science and Technology of China, Hefei,
Anhui 230026, P.R. China}
\affiliation{School of Astronomy, School of Physical Sciences,
University of Science and Technology of China, Hefei 230026, P.R. China}

\author{Spyros Basilakos}
\affiliation{National Observatory of Athens, Lofos Nymfon, 11852 Athens, Greece}
\affiliation{Academy of Athens, Research Center for Astronomy and Applied 
Mathematics,\\
Soranou Efesiou 4, 11527, Athens, Greece}
\affiliation{
School of Sciences, European University Cyprus, Diogenes Street, Engomi 1516 
Nicosia}
 
\author{Kuralay Yesmakhanova}
\affiliation{Ratbay Myrzakulov Eurasian International Centre for Theoretical 
Physics, Nur-Sultan 010009,  Kazakhstan}
 \affiliation{Eurasian National University, Nur-Sultan Astana 010008, 
Kazakhstan}

\begin{abstract}
We confront  $f(T,T_G)$ gravity, with Big Bang Nucleosynthesis (BBN)  
requirements. The former is obtained using both the torsion scalar,  as well 
as the teleparallel equivalent of the Gauss-Bonnet term, in the Lagrangian, 
resulting to modified Friedmann equations in which the extra torsional terms 
constitute an effective dark energy sector.  We   calculate the deviations of 
the freeze-out temperature $T_f$, caused by the extra torsion terms in  
comparison  to $\Lambda$CDM  paradigm. Then we impose five specific $f(T,T_G)$ 
models and we extract the constraints on the model parameters in order for the 
ratio $|\Delta T_f/ T_f|$ to satisfy the observational BBN bound. As we find,
in most of the  models the involved parameters are bounded in a narrow window 
around their General Relativity values as expected, as in the power-law 
model where the exponent   $n$ needs to be 
$n\lesssim 0.5$. Nevertheless the 
logarithmic model can easily satisfy the BBN constraints for large regions of 
the model parameters. This feature should be taken into account in future model 
building.

\end{abstract}

  
\maketitle

\section{Introduction}

There are two motivations that lead to  the construction of
modifications of gravity. The first is purely theoretical, namely to construct 
gravitational theories that do not suffer from the renormalizability problems 
of general relativity and thus being closer to a quantum description 
\cite{Stelle:1976gc,Addazi:2021xuf}. The second is cosmological, namely to 
construct gravitational theories that at a cosmological framework can 
describe the early and late accelarating eras 
\cite{Nojiri:2010wj,Clifton:2011jh,Nojiri:2017ncd,CANTATA:2021ktz,Ishak:2018his}
, as well as 
to alleviate various observational tensions \cite{Abdalla:2022yfr}.

There is a rich literature on modified and extended theories of gravity. One 
may start from the Einstein-Hilbert Lagrangian and add extra terms, 
resulting in    $f(R)$ gravity 
\cite{Starobinsky:1980te,Capozziello:2002rd,DeFelice:2010aj}, in  $f(G)$
gravity \cite{Antoniadis:1993jc,Nojiri:2005jg,DeFelice:2008wz}, in 
$f(G,{\cal{T}})$ theories \cite{Yousaf:2021xex}, in  
$f(P)$ 
gravity 
\cite{Erices:2019mkd,Marciu:2020ysf,Jimenez:2020gbw}
in Lovelock gravity \cite{Lovelock:1971yv, Deruelle:1989fj}, in Weyl gravity
\cite{Mannheim:1988dj}, in Horndeski/Galileon scalar-tensor theories 
\cite{Horndeski:1974wa,Deffayet:2009wt}  etc. Nevertheless, one can follow a 
different approach, and add new terms to the  equivalent torsional 
formulation of gravity, resulting to  $f(T)$ gravity 
\cite{Bengochea:2008gz,Cai:2015emx}, to $f(T,T_{G})$ gravity 
\cite{Kofinas:2014owa,Kofinas:2014aka, Kofinas:2014daa}, to $f(T,B)$ gravity 
\cite{Bahamonde:2015zma,Bahamonde:2016grb},
 to scalar-torsion
theories \cite{Geng:2011aj} etc. Torsional gravity has been proven to exhibit 
interesting  phenomenology, both at the cosmological framework  
\cite{ Chen:2010va, Wu:2010mn,Dent:2010nbw,Myrzakulov:2010vz,Zheng:2010am, 
Tamanini:2012hg, 
Bamba:2012vg, Dong:2012en, Karami:2012fu,Liu:2012fk,
Otalora:2013tba,Ong:2013qja,Chen:2014qtl, Farrugia:2016qqe,  Bejarano:2017akj,
Hohmann:2017jao,Bahamonde:2017wwk, 
Abedi:2018lkr,Golovnev:2018wbh,Krssak:2018ywd, Deng:2018ncg,Cai:2019bdh,
  Caruana:2020szx,Ren:2021tfi,Briffa:2021nxg, 
Benisty:2021sul, Dialektopoulos:2021ryi, Papagiannopoulos:2022ohv, 
    Papanikolaou:2022hkg},
 as well as at 
the level of local, spherically symmetric solutions   
\cite{Wang:2011xf,Boehmer:2011gw,Ferraro:2011ks,  Meng:2011ne,Rodrigues:2012qua,
 Rodrigues:2013ifa,Nashed:2013bfa,Bejarano:2014bca,Das:2015gwa, Mai:2017riq, 
Mustafa:2019eet, Nashed:2020kjh,   
Pfeifer:2021njm, Ren:2021uqb,
 Bahamonde:2021srr, 
Bahamonde:2022lvh,Huang:2022slc,Zhao:2022gxl}.

One crucial test that every modification of gravity should pass, that is 
usually underestimated in the literature, is the confrontation with the   Big 
Bang Nucleosynthesis (BBN)  
data\cite{Bernstein:1988ad,Kolb:1990vq,Olive:1999ij,Cyburt:2015mya,
Asimakis:2021yct}.  Specifically, the amount of 
modification needed in order to fulfill the late-time cosmological requirements 
must not at the same time spoil the successes of early-time cosmology, and 
among 
them the BBN phase. Hence, whatever are the advantages of a specific modified 
theory of gravity, if it cannot satisfy  the BBN constraints it must be 
excluded 
\cite{Torres:1997sn,Lambiase:2005kb,Lambiase:2011zz, 
 Anagnostopoulos:2022gej}.

In the present manuscript we are interested in investigating the BBN epoch in a 
universe governed by  $f(T,T_{G})$ gravity.  In particular, we desire to study 
various specific models that are known to lead to viable phenomenology, and 
extract constrains on the involved model parameters. 
The plan of the article is the following: In Section \ref{fTTGmpdel} we 
  briefly present $f(T,T_{G})$ gravity, extracting the field equations and 
applying them to a cosmological framework.  In Section 
\ref{BBNconss} we summarize the BBN formalism and we provide the difference in 
the freeze-out temperature caused by the   extra torsion terms. Then in Section
\ref{resultss} we investigate five specific $f(T,T_{G})$ models, confronting  
them with the observational BBN bounds. 
Finally, Section \ref{Conclusions} is devoted to the Conclusions.

\section{$f(T,T_{G})$ gravity } 
\label{fTTGmpdel}

In this section we briefly review   $f(T,T_G)$  gravity
\cite{Kofinas:2014owa,Kofinas:2014aka, Kofinas:2014daa}. As usual in torsional 
formulation of gravity we use   the tetrad  field as the dynamical variable,
which forms an orthonormal basis at the tangent space.  In a coordinate basis 
one can relate it  with the metric through $g_{\mu\nu}(x)=\eta_{AB}  
e^A_\mu (x)  e^B_\nu (x)$, where $\eta_{AB}=\text{diag}(-1,1,1,1)$, and with 
  Greek and Latin letters   denoting coordinate and tangent 
indices respectively.
Applying the  Weitzenb\"{o}ck connection  
$ {W}^\lambda_{\nu\mu}\equiv e^\lambda_A\: \partial_\mu e^A_\nu$ 
\cite{Cai:2015emx},  the corresponding torsion tensor is
\begin{equation}
\label{torsten}
{T}^\lambda_{\:\mu\nu} \equiv {W}^\lambda_{\nu\mu} - {W}^\lambda_{\mu\nu} = 
e^\lambda_A \: 
(\partial_\mu e^A_\nu - \partial_\nu e^A_\mu ) ~,
\end{equation}  
and then the torsion scalar is obtained through the contractions
\begin{equation}
\label{torsiscal}
T\equiv\frac{1}{4} T^{\rho \mu \nu} T_{\rho \mu \nu} + \frac{1}{2}T^{\rho \mu 
\nu }T_{\nu \mu \rho } 
- T_{\rho \mu }^{\ \ \rho }T_{\ \ \ \nu }^{\nu\mu} ~,
\end{equation}
and
 incorporates all information of the gravitational field. Used as a Lagrangian, 
the torsion scalar gives rise to exactly the same equations with General 
Relativity, that is why the theory was named  teleparallel equivalent of 
general 
relativity (TEGR).
 
Similarly to curvature gravity, where one can construct higher-order invariants 
such as the Gauss-Bonnet one, in torsional gravity one may construct 
higher-order torsional invariants, too. In particular,  since the curvature 
(Ricci) scalar and the torsion scalar differ by a total derivative, in 
\cite{Kofinas:2014owa} the authors followed the same recipe and extracted a 
higher-order torsional invariant which differs from the  Gauss-Bonnet one by a 
boundary term, namely
  \begin{eqnarray}
&&
\!\!\!\!\!\!\!\!\!\!\!\!\!\!
T_G=\left(K^{\kappa}_{\,\,\,\varphi\pi}K^{\varphi\lambda}_
{\,\,\,\,\,
\,\, \rho }K^{\mu}_{\,\,\,\,\chi\sigma}
K^{\chi\nu}_{\,\,\,\,\,\,\,\tau}
-2K^{\kappa\!\lambda}_{\,\,\,\,\,\,\pi}K^{\mu}_{
\,\,\,\varphi\rho}
K^{\varphi}_{\,\,\,\chi\sigma}K^{\chi\nu}_{\,\,\,\,\,\,\tau}
\right.
\nonumber\\
&&
\!\!\!\!\!\!\!\!\!\!\!\!\!\!
\left. 
+2K^{\kappa\!\lambda}_{\,\,\,\,\,\,\pi}K^{\mu}_{
\,\,\,\,\varphi\rho}
K^{\varphi\nu}_{\,\,\,\,\,\,\chi}K^{\chi}_{\,\,\,\,
\sigma\tau}
+2K^{\kappa\!\lambda}_{\,\,\,\,\,\,\pi}K^{\mu}_{
\,\,\,\,\varphi\rho}
K^{\varphi\nu}_{\,\,\,\,\,\,\,\sigma,\tau}\right)
\delta^{\pi\rho\sigma\tau}_{\kappa \lambda \mu \nu},
\label{TG}
\end{eqnarray}
where $K^{\mu\nu}_{\:\:\:\:\rho}\equiv-\frac{1}{2}\Big(T^{\mu\nu}_{\:\:
\:\:\rho} - T^{\nu\mu}_{\:\:\:\:\rho}-T_{\rho}^{\:\:\:\:\mu\nu}\Big)$ is the 
contortion tensor and 
 the generalized
$\delta^{\pi\rho\sigma\tau}_{\kappa \lambda \mu \nu}$ denotes the determinant
of the Kronecker deltas. Note that similarly to the Gauss-Bonnet term, the 
teleparallel equivalent of the Gauss-Bonnet term  $T_G$ is also a topological 
invariant in four dimensions.

Using the above torsional invariants one can construct the     new class of 
$f(T,T_G)$   
gravitational modifications, characterized by the action \cite{Kofinas:2014owa}
\begin{eqnarray}
S =\frac{M_P^2}{2}\!\int d^{4}x\,e\,f(T,T_G)\,,
\label{fGBtelaction}
\end{eqnarray}
with $M_P^2$ the reduced Planck mass.
The general field equations of the above action can be found in 
\cite{Kofinas:2014owa}, where one can clearly see that the   theory is 
different from $f(R)$, $f(R,G)$ and $f(T)$ gravitational modifications, and 
thus it corresponds to a novel class of modified gravity.

In this work we are interested in  the cosmological 
applications of $f(T,T_G)$   gravity. Hence,  
we consider  a spatially flat  Friedmann-Robertson-Walker (FRW) metric of the 
form
\begin{equation}
ds^{2}=-dt^{2}+a^{2}(t)\delta_{ij}dx^{i}dx^{j}\,,
\label{metriccosmo}
\end{equation}
with $a(t)$  the scale factor, which corresponds to the
diagonal tetrad
\begin{equation}
\label{vierbeincosmo}
e^{A}_{\,\,\,\mu}=\text{diag}(1,a(t),a(t),a(t)).
\end{equation}
In this case,  the torsion scalar  (\ref{torsiscal}) and the teleparallel 
equivalent of the Gauss-Bonnet term  (\ref{TG}) become 
\begin{eqnarray}
\label{Tcosmo1}
 &&T=6H^2\\
 &&T_G= 24H^2\big(\dot{H}+H^2\big) ,
 \label{TGcosmo1}
\end{eqnarray}
with $H=\frac{\dot{a}}{a}$ the Hubble parameter and where dots denoting
derivatives with respect to $t$.

The general field equations for the FRW geometry are 
\cite{Kofinas:2014aka} 
\begin{eqnarray}
&&
\!\!\!\!\!\!\!\!\!\!\!\!\!\!
f-12H^{2}f_{T}-T_G f_{T_G}
+24H^{3}\dot{f_{T_G}}=2M_P^{-2}(\rho_r+\rho_m)
\label{Fr1}
\\
&&\!\!\!\!\!\!\!\!\!\!\!\!\!\!
f-4\big(3H^2+\dot{H}\big)f_T-4H\dot{f_T}-T_G
f_{T_G}\nonumber\\
&&  \!\!\!\!\!\!\! \! \!
+\frac{2}{3H}T_G\dot{f_{T_G}}+8H^2\ddot{f_{T_G}}
=-2 M_P^{-2}(p_r+ p_m)\,,
\label{Fr2}
\end{eqnarray}
with $\dot{f_{T}}=f_{TT}\dot{T}+f_{TT_{G}}\dot{T}_{G}$,
$\dot{f_{T_{G}}}=f_{TT_{G}}\dot{T}+f_{T_{G}T_{G}}\dot{T}_{G}$, and 
$\ddot{f_{T_{G}}}=f_{TTT_{G}}\dot{T}^{2}+2f_{TT_{G}T_{G}}\dot{T}
\dot{T}_{G}+f_{T_{G}T_{G}T_{G}}\dot{T}_{G}^{\,\,2}+
f_{TT_{G}}\ddot{T}+f_{T_{G}T_{G}}\ddot{T}_{G}$,
and where  $f_{TT}$, $f_{TT_{G}}$,... denote multiple partial 
differentiations  with respect to $T$ and $T_{G}$.  Note that in the above 
equations we have also introduced the 
 radiation and matter sectors, corresponding to   perfect fluids with  energy
densities $\rho_r$,$\rho_m$ and pressures $p_r$,$p_m$, respectively. Lastly, 
we mention that the above equations for 
 $f(T,T_G)=-T+\Lambda$ recover the TEGR and General Relativity equations, where 
$\Lambda$ is the cosmological constant.

As we can see, we can re-write   the Friedmann equations (\ref{Fr1}) and
(\ref{Fr2}) in the usual form
 \begin{eqnarray}
\label{Fr1b}
&&
\!\!\!\!\!\!\!\!\!\!\!\!
3 M_P^2 H^2 =  \left(\rho_r +\rho_m + \rho_{DE} \right)   \\
\label{Fr2b}
&&\!\!\!\!\!\!\!\!\!\!\!\!
-2 M_P^2 \dot{H} = \left(\rho_r +p_r+\rho_m 
+p_m+\rho_{DE}+p_{DE}\right),
\end{eqnarray}
where we have defined 
  the effective dark energy  density and pressure  as
\begin{equation}
\label{rhode}\!\!\!\!\! \! 
\rho_{DE}\equiv\frac{M_P^2}{2}\! \left( \! 6H^2
\! -\! f\! +\! 12H^{2}f_{T}\! +\! T_G f_{T_G}\! -\! 24H^{3}\dot{f_{T_G}}\! 
\right),
\end{equation}
\begin{eqnarray}
\label{pde}
&&\!\!\!\!\!\!\!\!\!\!\!\!\! \! \!  \!\!\!\!\! \! \!\!\! 
p_{DE}  \equiv   \frac{M_P^2}{2 }\left[ 
-2(2\dot{H} + 3H^2)+f - 4\big(\dot{H} + 3H^2\big)f_T\right.\nonumber\\
&&
\left.  \!\! \! \!\!
  -4H\dot{f_T}-T_G
f_{T_G}+\frac{2}{3H}T_G\dot{f_{T_G}}+8H^2\ddot{f_{T_G}}
\right],
\end{eqnarray}
of gravitational origin.

\section{Big Bang Nucleosynthesis  constraints } 
\label{BBNconss}

Big Bang Nucleosynthesis (BBN) was a process that took place during radiation 
era.  Let us 
first present the framework which  provides the BBN 
constraints  through standard cosmology 
\cite{Bernstein:1988ad,Kolb:1990vq,Olive:1999ij,
Cyburt:2015mya,Asimakis:2021yct}. The first Friedmann 
equation from Einstein-Hilbert action   can be written as
\begin{equation}
3H^2=M_P^{-2}\rho,
\end{equation}
where $\rho=\rho_r+\rho_m$.
In the radiation era the radiation sector dominates hence we can write
\begin{eqnarray}\label{a}
 H^2\approx\frac{M_{P}^{-2}}{3} 
 \rho_r\equiv H_{GR}^2.
 \label{Frrad1}
\end{eqnarray}
In addition it is known that
 the energy density of relativistic particles  is
 \begin{eqnarray}
 \label{Temptime}
{\displaystyle \rho_r=\frac{\pi^2}{30}g_* {T}^4},
\end{eqnarray}
where $g_*\sim 10$  the effective number of degrees of freedom    
and ${T}$   the temperature. Thus , if we combine (\ref{Frrad1}) with  
(\ref{Temptime}) we obtain
\begin{eqnarray}
 H(T) \approx    \left(\frac{4\pi^3 
g_*}{45}\right)^{1/2}\frac{{T}^2}{M_{Pl} 
},
\label{HTemprel}
\end{eqnarray}
   where  $M_{Pl}= (8\pi)^{\frac{1}{2}} M_P = 1.22 \times 10^{19}$ GeV is the 
Planck mass.  

During the  radiation era the scale factor evolves as $a(t)\sim 
t^{1/2}$. Therefore, using the relation of Hubble parameter with scale 
factor 
we find that in the radiation era  the Hubble parameter evolves as  
$H(t)=\frac{1}{2t}$.  Combining the last one with (\ref{HTemprel}) we find 
the relation between temperature and time. Thus, we have ${\displaystyle 
\frac{1}{t}\simeq 
\left(\frac{32\pi^3 
g_*}{90}\right)^{1/2}\frac{{T}^2}{M_{Pl}}
}$
(or
${T}(t)\simeq (t/\text{sec})^{-1/2} $~MeV).

During the BBN  we have interactions between particles. For example we have 
interactions between  neutrons, protons, electrons and neutrinos, namely
$n+\nu_e\to p+e^-$, $n+e^+\to p+{\bar
\nu}_e$ and $n\to p+e^- +
{\bar \nu}_e$. We name the conversion rate from a particle A to particle B as 
$\lambda_{BA}$. Hence, the conversion rate from neutrons to protons is 
$\lambda_{pn}$ and it is equal to the sum of the three interaction 
conversion rates   written above. Therefore, the calculation of the 
neutron 
abundance 
arises from the protons-neutron
conversion rate \cite{Olive:1999ij,Cyburt:2015mya}
 \begin{equation}
 \lambda_{pn}({T})=\lambda_{(n+\nu_e\to p+e^-)}+\lambda_{(n+e^+\to p+{\bar
\nu}_e)}+\lambda_{(n\to p+e^- +
{\bar \nu}_e)}\,
 \end{equation}
and its inverse $\lambda_{np}({T})$, and therefore for the  total 
rate we have
$    \lambda_{tot}({T})=\lambda_{np}({T})+\lambda_{pn}({T})$.
Now, we assume that the various particles (neutrinos, electrons, photons) 
temperatures   are the same, and low enough in order to use 
the Boltzmann distribution instead of the
Fermi-Dirac one,  and we neglect  the electron mass  
 compared to the electron and neutrino energies. The final expression for the 
conversion rate is
\cite{Torres:1997sn,Lambiase:2005kb,Lambiase:2011zz, 
 Anagnostopoulos:2022gej}
 \begin{equation}\label{Lambdafin}
    \lambda_{tot}({T}) =4 A\, {T}^3(4! {T}^2+2\times 3! {  Q}{T}+2!
{  Q}^2)\,,
 \end{equation}
where  ${
Q}=m_n-m_p=1.29 \times10^{-3}$GeV is   the   mass difference between neutron 
and proton 
and $A=1.02 \times 10^{-11}$~GeV$^{-4}$.

We proceed in calculating the corresponding freeze-out temperature. This will 
arise 
comparing the universe expansion rate $\frac{1}{H}$ with $ 
\lambda_{tot}\left(T\right)$. In particular, if $\frac{1}{H} \ll
\lambda_{tot}\left(T\right)$, namely if the expansion time is much smaller than
the interaction time, we can consider thermal equilibrium 
\cite{Bernstein:1988ad,Kolb:1990vq}. On the contrary, if $\frac{1}{H}
\gg\lambda_{tot}\left(T\right)$ then particles 
 do not have enough   time   to interact so they 
 decouple. The freeze-out 
temperature $T_{f}$, in which the decoupling takes place, corresponds to 
 $H (T_{f})= \lambda_{tot}\left(T_{f}\right) \simeq c_q \,T_f^5$, with $c_q 
\equiv 4A \, 4! \simeq 9.8 \times 10^{-10} \, {\rm GeV}^{-4}$
\cite{Torres:1997sn,Lambiase:2005kb,Lambiase:2011zz, 
 Anagnostopoulos:2022gej}.  
Now if we use (\ref{HTemprel}) and $H (T_{f})= \lambda_{tot}\left(T_{f}\right) 
\simeq c_q \,T_f^5$, we acquire
\begin{equation}
T_{f}=\left(\frac{4\pi ^3 g_{*}}{45M_{Pl}^2c_{q}^2}\right)^{1/6}\sim 0.0006 
~{\rm 
GeV}.
\end{equation}

Using modified theories we obtain extra terms in energy density due to the 
modification of gravity. The  first Friedmann equation (\ref{Fr1b})
  during radiation era   becomes
\begin{equation}\label{b}
3M_P^{2}H^2= \rho_r+\rho_{DE} ,
\end{equation}
where $\rho_{DE}$ must be very small compared to $\rho_r$ in order to be in 
accordance with observations. Hence, we can write (\ref{b}) using 
(\ref{a}) as
\begin{equation}
H=H_{GR}\sqrt{1+\frac{\rho_{DE}}{\rho_{r}}}=H_{GR}+\delta H,
\end{equation}
where $H_{GR}$ is the Hubble parameter of standard cosmology.  
Thus, we have
$
\Delta  H=\left(\sqrt{1+\frac{\rho_{DE}}{\rho_{r}}}-1\right)H_{GR}$,
which quantifies    the deviation  from standard cosmology, i.e form  
$H_{GR}$. This will 
lead to a 
deviation in the  freeze-out 
temperature  $\Delta 
T_{f}$. Since   $H_{GR} = \lambda_{tot}\approx c_q\,T_f^5$ and 
$\sqrt{1+\frac{\rho_{DE}}{\rho_r}}\approx 
1+\frac{1}{2}\frac{\rho_{DE}}{\rho_r}$, we easily find
\begin{equation}
\left(\sqrt{1+\frac{\rho_{DE}}{\rho_{r}}}-1\right)H_{GR}=5c_q\,T_{f}^4\Delta  
T_{f},
\end{equation}
and finally 
\begin{equation}
\frac{\Delta  
T_{f}}{T_{f}}\simeq\frac{\rho_{DE}}{\rho_{r}}\frac{H_{GR}}{10c_q\,T_{f}^5},
\label{finalexpress}
\end{equation}
 where we used that  $\rho_{DE}<<\rho_{r}$ during BBN era. 
This theoretically calculated $\frac{\Delta  
T_{f}}{T_{f}}$ should be compared with the observational bound
 \begin{equation}
 \label{deltaTbound}
    \left|\frac{\Delta {T}_f}{{T}_f}\right| < 4.7 \times 10^{-4}\,,
 \end{equation}
 which is obtained from the observational estimations of
the baryon mass
fraction   
converted to ${}^4 He$ 
\cite{Coc:2003ce,Olive:1996zu,Izotov:1998mj,Fields:1998gv,Izotov:1999wa,
Kirkman:2003uv, Izotov:2003xn}.

\section{BBN constraints on $f(T,T_{G})$ gravity  }
\label{resultss}

In this section we will apply the BBN analysis in the case of $f(T,T_{G})$ 
gravity. Let us mention here that in general, in modified gravity, 
inflation is  not   straightaway driven by an inflaton field but the 
inflaton is hidden inside the gravitational modification, i.e. it is one of the 
extra scalar degrees of freedom of the modified graviton. Hence, in such 
frameworks reheating is usually performed gravitationally, and the reheating 
and BBN temperatures may differ from  standard ones. Nevertheless,   in the 
present work we make the assumption that we do not   deviate 
significantly from the successful concordance scenario, in order to examine 
whether $f(T,T_G)$  gravity can at first pass BBN constraints or not. Clearly a 
more general analysis should be performed in a separate project, to cover more 
radical cases too.  In the following, we will examine five 
specific models that are  considered to be viable in the literature.

\subsubsection{ Model I:  $f=-T+\beta_1\sqrt{T^2+\beta_2T_G}$ }

Firstly we investigate the model $f=-T+\beta_1\sqrt{T^2+\beta_2T_G}$ 
\cite{Kofinas:2014daa}.
Since in  our analysis we focus on the radiation era where the Hubble parameter 
$H(t)=\frac{1}{2t}$, we can express the 
 derivatives of the Hubble parameter 
as powers of the Hubble parameter itself, e.g. $\dot{H}=-2H^2$ and 
$\ddot{H}=8H^3$. Additionally, in order to eliminate one model parameter we 
will apply the Friedmann equation at present time, requiring
\begin{equation}\label{c}
\Omega_{DE0}\equiv 
\rho_{DE0}/(3M_P^{2}H_0^2),
\end{equation}
where $\Omega_{DE}$ is the dark energy density parameter and with the subscript 
``0'' denoting the value of a quantity at present time. Doing so, and inserting 
 $f=-T+\beta_1\sqrt{T^2+\beta_2T_G}$ into (\ref{rhode}) and then into 
(\ref{finalexpress}),
we finally find
\begin{multline}\label{model1}
\!\!\!\!\!\!\!\!
\frac{\Delta T_f}{T_f}=(10c_qT_f^3)^{-1}\zeta 
H_0\Omega_{DE0}\left(3-2\beta_2\right)^{-3/2}\\
\cdot\left(9-15\beta_2+6\beta_2^2\right)\left[
\left(3+2\beta_2\right)H_0^2+2\beta_2\dot{H}_0\right]^{3/2}\\\cdot
\left[\left(9+3\beta_2-2\beta_2^2\right)H_0^4+9\beta_2H_0^2\dot{H}
_0+\beta_2^2H_0\ddot{H}_0\right]^{-1},
\end{multline}
where
\begin{align}\label{defzeta}
\zeta\equiv\left(\frac{4\pi^3g_{*}}{45}\right)^{\frac{1}{2}}M_{Pl.}^{-1}~.
\end{align}
In this expression we insert \cite{Planck} 
\begin{align}\label{deo}
\Omega_{DE0} \approx0.7, \quad 
 H_{0}=1.4\times 
10^{-42} ~{\rm GeV}, 
\end{align}
and   the derivatives of the Hubble 
function at present are calculated through 
 $\dot{H}_{0}=-H_{0}^2\left(1+q_{0}\right)$ and 
$\ddot{H}_{0}=H_{0}^3\left(j_{0}+3q_{0}+2\right)$
 with $q_{0}=-0.503$ the current decceleration parameter of the 
Universe~\cite{Planck}, and $j_{0}=1.011$ the current jerk parameter 
\cite{Visser:2003vq,Mamon:2018dxf}. Hence,  $\dot{H}_{0}\approx -9.7\times 
10^{-85}\, $GeV$^2$ and $\ddot{H}_{0}\approx 4.1\times 
10^{-126}\,$GeV$^3$.

Using the BBN constraint (\ref{deltaTbound})
   we conclude that 
 $\beta_2  \in\left(-2.98,-2.93\right)$
$\cup$ $\left(0.99,1.01\right)$, 
where we have used (\ref{c}) to find
\begin{multline}\!\!\!\!\!
\beta_1=\sqrt{3}H_0\Omega_{DE0}\left[
\left(3+2\beta_2\right)H_0^2+2\beta_2\dot{H}_0\right]^{3/2}\\ \!\!
\cdot\left[
\left(9+3\beta_2-2\beta_2^2\right)H_0^4+9\beta_2H_0^2\dot{H}_0+\beta_2^2H_0\ddot
{H}_0\right]^{-1}.
\end{multline}
Using the above range of $\beta_2$ we find that
$\beta_1  \in\left(2.09\times 10^{-26},0.001\right)$
$\cup$ $\left(1.380,1.384\right)$.

\begin{figure}[ht] 
\centering
\includegraphics[angle=0,width=0.49\textwidth]{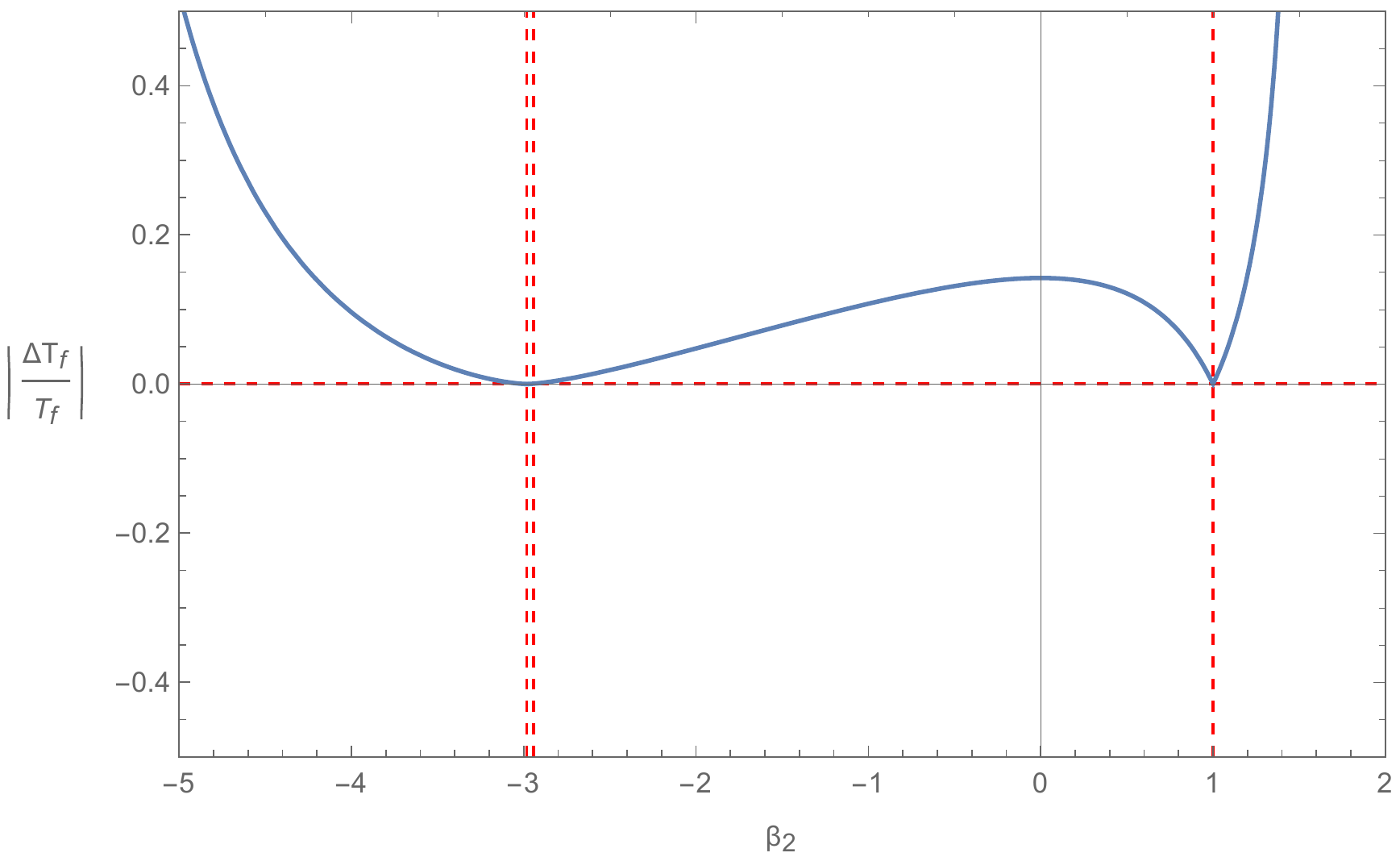}
 \caption{{\it{   $|\Delta T_f/  T_f|$ 
vs the model parameter $\beta_2$   (blue solid  curve), for Model I:  
$f=-T+\beta_1\sqrt{T^2+\beta_2T_G}$. The allowed range of $\beta_2$, 
 where (\ref{deltaTbound}) is satisfied (horizontal red dashed line),  is 
within the vertical dashed lines. }} }
\label{fig1}
\end{figure}

In Fig. \ref{fig1} we depict $|\Delta T_f/  T_f|$ 
appearing in 
(\ref{model1})
 versus the model parameter $\beta_2$. As we can see the allowed range is 
within the vertical dashed lines.  

\subsubsection{ Model II:  $f=-T+a_1T^2+a_2T\sqrt{|T_G|}$ }

Let us now study the case  $f=-T+a_1T^2+a_2T\sqrt{|T_G|}$, where $a_1$,$a_2$ 
are the free parameters of the theory \cite{Kofinas:2014daa}.
In this case we find
\begin{multline}\label{1}
\frac{\Delta T_f}{T_f}=
\frac{3}{10}c_q^{-1}\zeta^3T_f\Bigg\{\frac{\Omega_{DE0}}{
3H_0^2}\\
-\sqrt
{6}a_2\Bigg[\frac{\sqrt{H_0^2+\dot{H_0}}}{6H_0} 
 \Big(6-\frac{2\dot{H}_0^2-H_0\ddot{H}_0}{H_0^2+\dot{H}
_0}\Big)-1\Bigg]\Bigg\}.
\end{multline}

Using the constraint (\ref{deltaTbound}), ans since  according to (\ref{1})  
$\Delta {  T}_f/{  T}_f$ is 
linear in $a_2$, we deduce that 
(\ref{1}) is valid for a small region around   $ 
2.7\times 10^{83}~{\rm GeV}^{-2}$, where we have used the constraint from 
current cosmological era (\ref{c})
\begin{multline}\label{model II}
a_1=\frac{\Omega_{DE0}}{18H_0^2}-\sqrt{6}a_2\frac{\sqrt{H_0^2+\dot{H_0}}}{36H_0}
\left[6-\frac{2\dot{H}_0^2-H_0\ddot{H}_0}{\left(H_0^2+\dot{H}_0\right)^2}\right]
.
\end{multline}
Using the above value of $a_2$ we find that
$a_1 =-1.1\times 10^{83}~{\rm GeV}^{-2}$.

\subsubsection{ Model III: 
$f\!=\!-T+\beta_1\sqrt{T^2\!+\!\beta_2T_G}+a_1T^2+a_2T\sqrt{|T_G|}$ }

Now we analyze the  model 
$f=-T+\beta_1\sqrt{T^2+\beta_2T_G}+a_1T^2+a_2T\sqrt{|T_G|}$, where we have four 
free parameters, namely $\beta_1$, $\beta_2$, $a_1$, $a_2$ 
\cite{Kofinas:2014daa}. In order to simplify the analysis we will impose the 
constraint   $-2.99<\beta_2<\frac{3}{2}$, obtained above.

In this case we find
\begin{multline}\label{model III}
\frac{\Delta 
T_f}{T_f}=-\left(60c_qT_f^3\right)^{-1}\!\left\{3\sqrt{12}
\beta_1\left(3\!-\!2\beta_2\right)^{-1/2}
\left(1\!+\!\beta_2\!-\!2\beta_1\right)\right. \\
 -18\Big\{\frac{\Omega_{DE0}}{3H_0^2}+\frac{\sqrt{12}\beta_1}{18H_0^3}
\left 
[\left(3+2\beta_2\right)H_0^2+2\beta_2\dot{H}_0\right]^{-1/2} \\
\left.\left. \!\!\!\!\!\!\!\!\!\!\!\!\!\!\!\!\!\!\!\!\!\!\!\!
\cdot\left[\left(3\!-\!6\beta_1\!+\!2\beta_2\right)H_0^2+2\beta_2\dot{H}_0\right
] \right.\right.\\
\left.\left.\ \ \ \ \ \ \ \ -
\frac{a_2}{\sqrt{6}H_0}\sqrt{H_0^2+\dot{H}_0}
\Bigg[6-\frac{2\dot{H}_0^2-H_0\ddot{H}_0}{ (H_0^2+\dot{H}_0 )^2}
\Bigg]+\sqrt{6}a_2\right.\right.\\
\left.\left.-\frac{\sqrt{12}\beta_1\beta_2}{18H_0^3}\left[
\left(3+2\beta_2\right)H_0^2+2\beta_2\dot{H}_0\right]^{-3/2}\right.\right.\\
\!\!\!\!\!\!\!\!
 \cdot
\left[
\left(3+2\beta_2\right)H_0^4+\left(9+8\beta_2\right)H_0^2\dot{H}
_0 \right.\nonumber\\
\left.\left.
+\beta_2\left(4\dot{H}_0^2+H_0\ddot{H}_0\right)\right]   
\Big\}\zeta^2T_f^4\right\}\zeta.
\end{multline}
Observing that expression
(\ref{model III}) is 
linear in $a_2$, and using the constraint (\ref{deltaTbound}) and two values 
for $\beta_1$ from the aforementioned range we extracted in model I,  i.e. 
$\beta_1=1.4$ and $\beta_2=1$,  we find that (\ref{1}) is valid for a small 
region around the point $ -3.5\times 10^{83}~{\rm GeV}^{-2}$. Using another set 
of values ($\beta_1=0.001$, $\beta_2\approx-2.96$) we find that  (\ref{1}) is 
valid for a small region around the point $ -5.3\times 10^{83}~{\rm GeV}^{-2}$, 
where we have used 
\begin{multline}\!\!\!\!\!\!\!
a_1=\frac{\Omega_{DE0}}{18H_0^2}+\frac{\sqrt{12}\beta_1}{108H_0^3}\left[
\left(3+2\beta_2\right)H_0^2+2\beta_2\dot{H}_0\right]^{-1/2}\\
\left[\left(3-6\beta_1+2\beta_2\right)H_0^2+2\beta_2\dot{H}_0\right]\\
-\frac{\sqrt{6}}{36}\frac{a_2}{H_0}\sqrt{H_0^2+\dot{H}_0}\left(6-\frac{2\dot{H}
_0^2-H_0\ddot{H}_0}{\left(H_0^2+\dot{H}_0\right)^2}\right)\\
-\frac{\sqrt{12}\beta_1\beta_2}{108H_0^3}\left[
\left(3+2\beta_2\right)H_0^2+2\beta_2\dot{H}_0\right]^{-3/2}\\
\times\left[\left(3+2\beta_2\right)H_0^4+\left(9+8\beta_2\right)H_0^2\dot{H}
_0+\beta_2\left(4\dot{H}_0^2+H_0\ddot{H}_0\right)\right],
\end{multline}
from (\ref{c}).
Imposing the above range of $a_2$ we find that
$a_1 =
1.4\times 10^{83}~{\rm GeV}^{-2}$ for the first case and $a_1= 
2.2\times 10^{83}~{\rm GeV}^{-2}$ for the second.

\subsubsection{ Model IV: $f=-T+\beta_1\left(T^2+\beta_2T_G\right)^n$}

As a  next model we consider   the power-law model 
$f=-T+\beta_1\left(T^2+\beta_2T_G\right)^n$, where the free parameters are   
$\beta_1$, $\beta_2$, $n$. In this model we use values of $\beta_1$, $\beta_2$ 
in order to constrain the power $n$.
In this case, repeating the above steps, we find
\begin{eqnarray}
\label{model IV}
&&\!\!\!
\frac{\Delta 
T_f}{T_f}=\left(10c_q\right)^{-1}\Omega_{DE0}H_0^{2\left(1-n\right)}\zeta^{4n-1}
T_f^{8n-7}\left(3-2\beta_2\right)^{n-2}\nonumber\\
&&\cdot
\left[
\left(3+2\beta_2\right)H_0^2+2\beta_2\dot{H}_0\right]^{2-n}
\left[\left(9-12\beta_2+4\beta_2^2\right)\right. \nonumber\\
&&\left.\ \ \
-2n\left(18-39\beta_2+18\beta_2^2\right)+
16n^2\beta_2
\left(2\beta_2-3\right)\right]\nonumber\\
&&\cdot
\left\{
\left(9+12\beta_2+4\beta_2^2\right)H_0^4+4\beta_2\left(3+2\beta_2\right)H_0^2
\dot{H}
_0+4\beta_2^2\dot{H}_0^2\right.\nonumber\\
&&
\left.-2n\left[
\left(18+15\beta_2+2\beta_2^2\right)H_0^4+\beta_2\left(27+12\beta_2\right)H_0^2
\dot{H}_0\right.\right.\nonumber\\
&&
\left.\left.+6\beta_2^2\dot{H}_0^2+2\beta_2^2H_0\ddot{H}_0\right]
+2n^2\beta_2\left[4\left(3+2\beta_2\right)H_0^2\dot{H}
_0\right.\right.\nonumber\\
&&
\left.\left.+4\beta_2\dot{H}_0^2+2\beta_2H_0\ddot{H}_0\right]\right\}^{-1}.
\end{eqnarray}

We use the constraint (\ref{deltaTbound}) and four values for $\beta_2$  from 
the range we extracted in model I above. For $\beta_2\approx -2.9$ we find that 
 the constraint (\ref{deltaTbound}) is valid for
$n \lesssim 0.5$. Similarly, using the value $\beta_2=-2$   we find $n 
\lesssim 0.47$, while  for $\beta_2=-1$  we find 
$n \lesssim 0.46$. Finally,  for $\beta_2=1$ we find $n 
\lesssim 0.47$. We mention  that we have used the relation
\begin{multline}
\!\!\!\!\!\!\!
\beta_1=-6\left(12\right)^{-n}\!H_0^{2(1-n)}\Omega_{DE0}\!\left[\!
\left(3\!+\!2\beta_2\right)H_0^2\!+\!2\beta_2\dot{H}_0\right]^{2-n}\\
\cdot
\left\{\!
\left(9+12\beta_2+4\beta_2^2\right)H_0^4+4\beta_2\left(3+2\beta_2\right)H_0^2\do
t{H}
_0+4\beta_2^2\dot{H}_0^2\right.\\
\left.-2n\left[
\left(18+15\beta_2+2\beta_2^2\right)H_0^4+\beta_2\left(27+12\beta_2\right)H_0^2
\dot{H}_0\right.\right.\\
\left.\left. \!\!\! \!+6\beta_2^2\dot{H}_0^2+2\beta_2^2H_0\ddot{H}_0\right]
+2n^2\beta_2\left[4\left(3+2\beta_2\right)H_0^2\dot{H}_0\right.\right.\\
\left.\left.
\ \ \ \ \ \ \ \ \ \ \ \ \ \ \ \ \ \ \ \ \ \ \ \ \ \ \ \ 
+4\beta_2\dot{H}_0^2+2\beta_2H_0\ddot{H}_0\right]\right\}^{-1},
\end{multline}
which arises from  (\ref{c}).

Now taking $\beta_2\approx -2.9$, $n \lesssim 0.5$ we find 
$\beta_1\in\left[ -6.1\times 10^{-82},0.0007\right]~{\rm 
GeV}^{2\left(1-2n\right)}$. Similarly, for $\beta_2=-2$, $n \lesssim 0.47$ we 
find $\beta_1\in\left[-3.5\times 10^{-74},5.9\times 10^{-6}\right]~{\rm 
GeV}^{2\left(1-2n\right)}$, while using $\beta_2=-1$, $n \lesssim 0.46$ we 
find $\beta_1\in\left[-4.4\times 10^{-58},1.2\times 10^{-6}\right]~{\rm 
GeV}^{2\left(1-2n\right)}$. Finally, for  $\beta_2=1$, $n \lesssim 0.47$ we 
find $\beta_1\in\left[-6.4\times 10^{-8},9.0\times 10^{-6}\right]~{\rm 
GeV}^{2\left(1-2n\right)}$.

\begin{figure}[ht] 
\centering
\includegraphics[angle=0,width=0.49\textwidth]{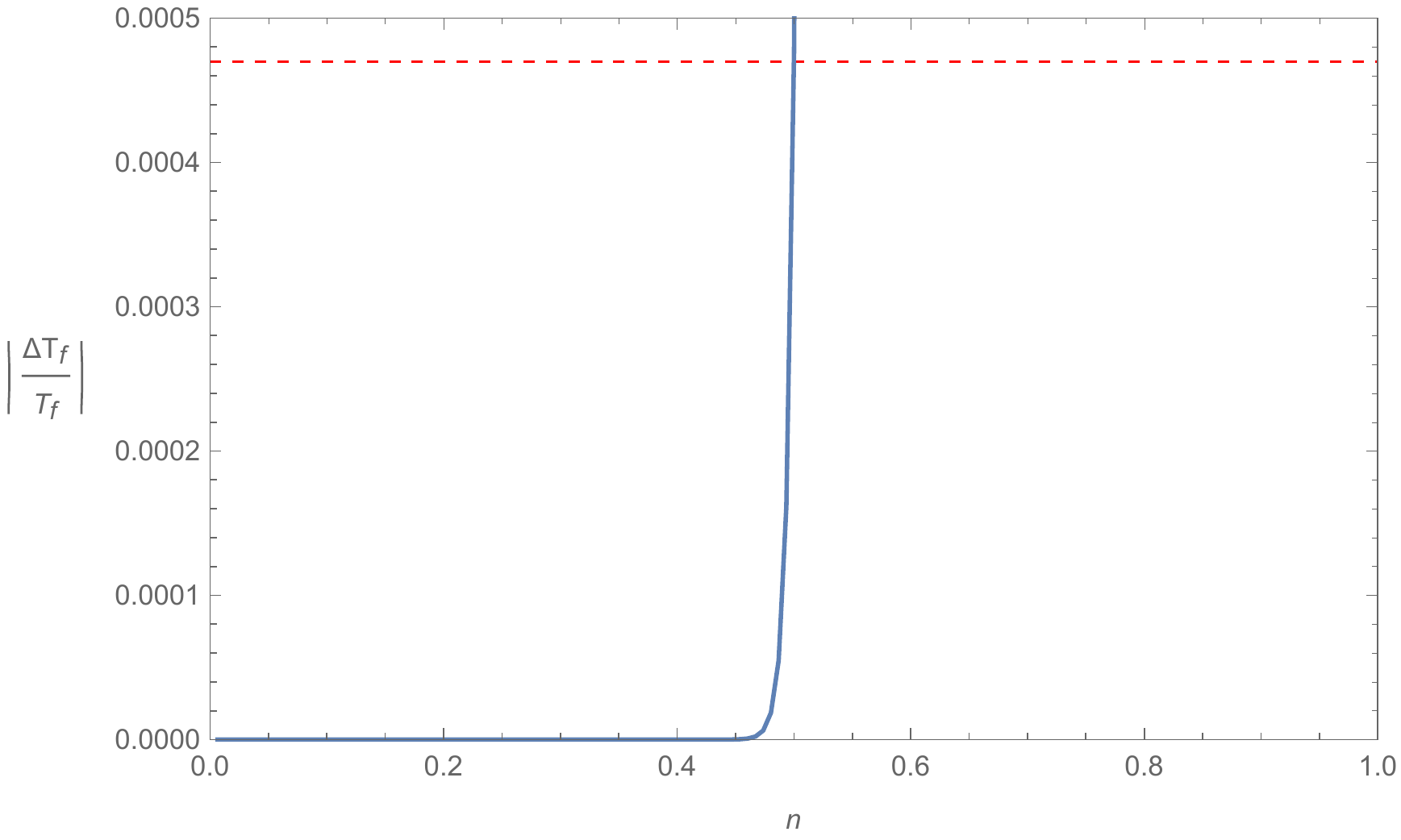}
\vspace{-0.2cm}\caption{{\textit{$|\Delta {T}_f/{  T}_f|$ 
vs the model parameter $n$   (blue solid  curve), for Model IV: 
$f=-T+\beta_1\left(T^2+\beta_2T_G\right)^n$   with $\beta_2\approx-2.90$,  and 
the upper bound for $|\Delta {T}_f/{  T}_f|$  from (\ref{deltaTbound}) (red 
dashed line).   As 
we observe,  constraints from BBN require $n\lesssim 0.5$.}} }
\label{fig4}
\end{figure}

In order to provide the above results in a more transparent way, 
in Fig. \ref{fig4},  we present  $|\Delta {T}_f/{  T}_f|$  from
(\ref{model IV})
in terms of  the model parameter $n$. As we observe,   $n$ needs to be 
$n\lesssim 0.5$ to pass the BBN constraint (\ref{deltaTbound}).

\subsubsection{Model V: $f=-T+\alpha\ln\beta_1\left(T^2+\beta_2T_G\right)^n$}

The last model we examine is  the logarithmic one, characterized by
$f=-T+\alpha\ln\beta_1\left(T^2+\beta_2T_G\right)^n$, where   $\beta_1$, 
$\beta_2$, $n$ are the free parameters. Repeating the above analysis we find
\begin{multline}\label{model V}
\!\!\!\!\!\!
\frac{\Delta T_f}{T_f}=\left(10c_q\zeta 
T_f^7\right)^{-1}H_0^2\Omega_{DE0}
\Big\{\ln\beta_1+n\left[\ln12\right. \\
\left.
\ \  
\left.+4\ln\left(\zeta 
T_f^2\right)+\ln\left(3\!-\!2\beta_2\right)
\right.\right.\\
 \left. \ \ \ \ \ \ \ \ \ \ \ \ \ \ \ \ 
-2\left(3\!-\!2\beta_2\right)^{-2}
 \left(18\!-\!39\beta_2\!+\!18\beta_2^2\right)\right]\Big\}\\ \!
\cdot
 \Big\{\ln\beta_1+n\left\{\ln12+2\ln(H_0)+\ln[
(3\!+\!2\beta_2)H_0^2\!+\!2\beta_2\dot{H}_0] \right.\\        
\left.\left. 
  -2\left[\left(3+2\beta_2\right)H_0^2+2\beta_2\dot{H}_0\right]^{-2} 
\left[\left(18+15\beta_2+2\beta_2^2\right)H_0^4\right.\right.\right.\\
 \left.\left.
  +\beta_2\left(27\!+\!12\beta_2\right)H_0^2\dot{H}_0\!+\!6\beta_2^2\dot
{H}_0^2\!+\!2\beta_2^2H_0\ddot{H}_0\right]\right\}\!\!\Big\}^{-1},
\end{multline}
where using   relation (\ref{c}) we find
\begin{multline}\!\!\!\!\!
\alpha=-6H_0^2\Omega_{DE0}\Big\{\ln\beta_1+n\Big\{
\ln12+2\ln(H_0) \\
\left.\left.+ \ln\left[\left(3\!+\!2\beta_2\right)H_0^2\!+\!2\beta_2\dot{H}
_0\right] \right.\right.\\        
\left.\left.
-2\left[(3\!+\!2\beta_2)H_0^2+2\beta_2\dot{H}_0\right]^{-2}
\Big[(18\!+\!15\beta_2\!+\!2\beta_2^2)H_0^4\right.\right. \\
+  \beta_2\left(27\!+\!12\beta_2\right)H_0^2\dot{H}_0+6\beta_2^2\dot{
H}_0^2+2\beta_2^2H_0\ddot{H}_0\Big]\Big\}\Big\}^{-1}.
\end{multline}

We consider the values $\beta_1=0.001~{\rm GeV}^{-4n}$, $\beta_2\approx 
-2.9$ 
and we find  that $n$ is allowed to take every value 
apart from   $-0.0003$ and a very small region around it, since 
(\ref{model V}) diverges. Moreover,  $\alpha$ is allowed to take every value 
apart from $0$,  which is the value it obtains using the above   
narrow window  for $n$. 
Using the same considerations as the above models,  we find that for 
$\beta_1=0.001~{\rm GeV}^{-4n}$, $\beta_2=-2$ the value of  $n$ is allowed to 
take every value apart from  $-0.012$ and $a$  every 
value but $0$. Similarly, for $\beta_1=0.001~{\rm GeV}^{-4n}$, 
$\beta_2=-1$ we find that $n\neq-0.018$  and $a\neq0$, while for  
$\beta_1=0.001~{\rm GeV}^{-4n}$, $\beta_2=1$ we find   $n\neq -0.018$  and 
$a\neq0$.

\begin{figure}[ht] 
\centering
\includegraphics[angle=0,width=0.49\textwidth]{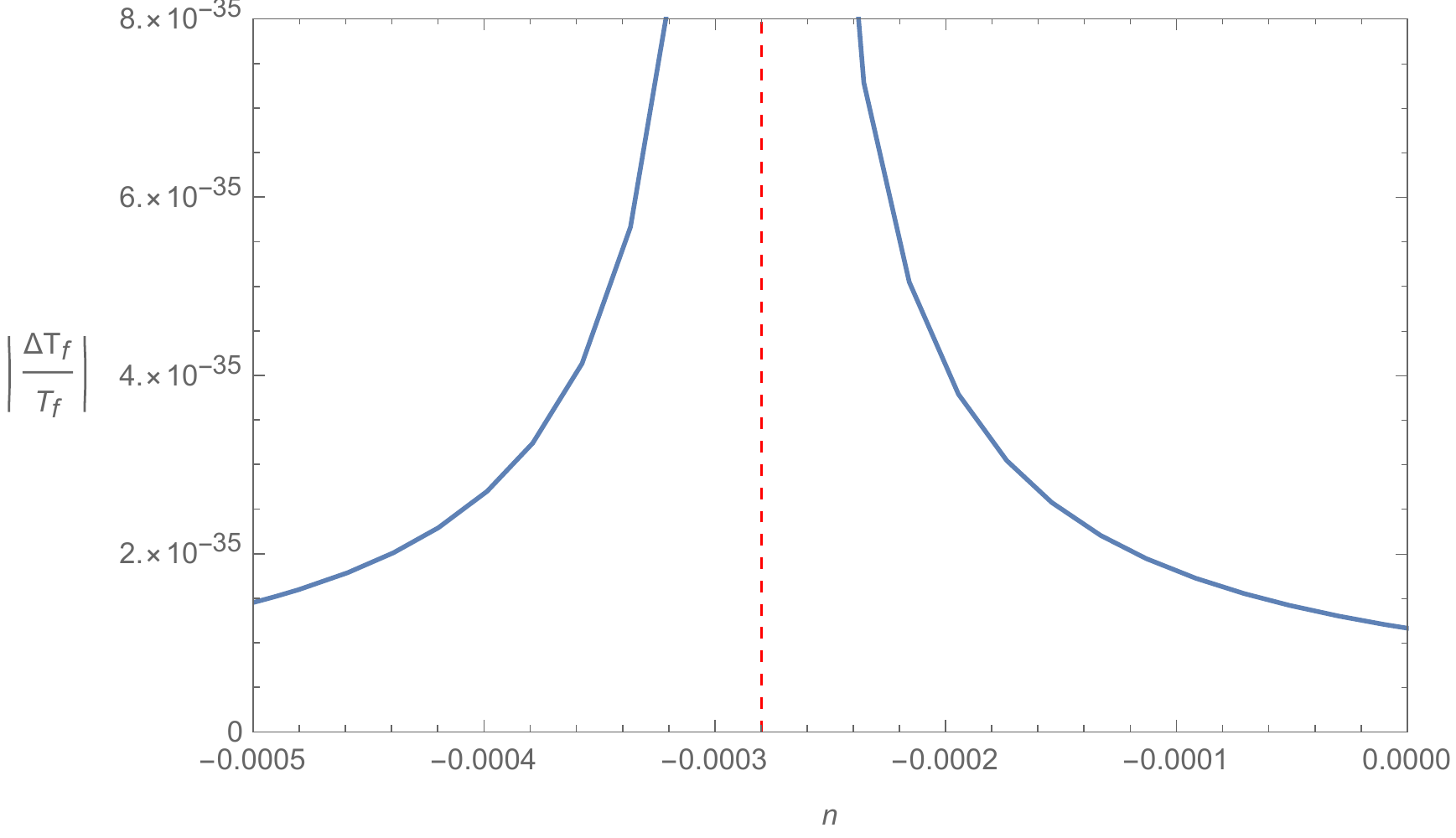}
\vspace{-0.2cm}\caption{{
\textit{  $|\Delta T_f/  T_f|$ 
vs the model parameter $n$   (blue solid  curve), for Model V: 
$f=-T+\alpha\ln\beta_1\left(T^2+\beta_2T_G\right)^n$, choosing 
$\beta_1=0.001~{\rm GeV}^{-4n}$, $\beta_2\approx-2.7$. The vertical dashed line 
at  $n-0.0003$ denotes the point where  (\ref{model V})  diverges.}} }
\label{fig5}
\end{figure}

As an example, in Fig. \ref{fig5}     we present    $|\Delta T_f/  T_f|$ 
from
(\ref{model V})
as a function of  the model parameter $n$. The model parameter $n$ is allowed 
to take all possible   values  except those values around 
a very small region centered  at     $-0.0003$,  in  which   
(\ref{model V}) diverges. Hence, we conclude that the logarithmic  $f(T,T_G)$ 
model can easily satisfy the BBN bounds.

\section{Conclusions}
\label{Conclusions}
 
Modified gravity aims to provide explanations for various epochs of the 
Universe evolution, and at the same time to improve the renormalizability 
issues of General Relativity. Nevertheless, despite the specific advantages at 
a given era of the cosmological evolution one should be very careful not to 
spoil other, well understood and significantly constrained, phases, such as the 
Big Bang Nucleosynthesis (BBN) one.

In particular, there are many modified gravity models, which are constructed 
phenomenologically in order to be able to describe the late-time universe 
evolution at both background and perturbation level. Typically, these models 
are confronted with observational data such as Supernovae Type Ia (SNIa),   
Baryonic Acoustic 
Oscillations (BAO), Cosmic Microwave Background (CMB),   Cosmic Chronometers 
(CC), Gamma-ray Bursts (GRB), 
growth data, etc. The problem is that although modified gravity scenarios, 
through the extra terms they induce, are very efficient in describing the 
late-time universe, quite often they induce significant terms at  early times 
too, and thus spoiling the early-time evolution, such as the BBN phase, in 
which the concordance cosmological paradigm is very successful. Hence, 
independently of the late-universe successes that a modified gravity model may 
have, one should always examine whether the model can pass the BBN constraints 
too.

In the present work we confronted one interesting class of gravitational 
modification, namely $f(T,T_G)$ gravity, with BBN requirements. The former is 
obtained using both the torsion scalar, as well as the teleparallel equivalent 
of the Gauss-Bonnet term, in the Lagrangian. Hence, one obtains modified 
Friedmann equations in which the extra torsional terms constitutes an effective 
dark energy sector.
 
 We started by  calculating the deviations of the freeze-out temperature 
$T_f$, caused by the extra torsion terms, in  comparison  to $\Lambda$CDM 
paradigm. We imposed five specific $f(T,T_G)$ models
 that have been proposed in the 
literature in phenomenological grounds, i.e. in order to be able to describe 
the late-time evolution and lead 
to acceleration without an explicit cosmological constant. Hence, we 
extracted the 
constraints on the model parameters in order for the ratio $|\Delta T_f/  
T_f|$ to satisfy the BBN bound  $ \left|\frac{\Delta {T}_f}{{T}_f}\right| 
< 4.7 \times 10^{-4}$. As we found, in most of the  models the involved 
parameters are bounded in a narrow window around their General Relativity 
values, as expected. However, the logarithmic model can easily satisfy the BBN 
constraints for large regions of the model parameters, which acts as an 
advantage for this scenario.

We stress here that we did not  fix the cosmological parameters to their 
General Relativity values, on the contrary we left them completely free and 
we examined which parameter regions are allowed if we want the models to  pass 
the BBN constraints. The fact that  in most models the parameter regions are 
constrained to a narrow window around their General Relativity values was 
in some sense expected, but 
in general is not guaranteed or known a priori, since many modified gravity 
models are completely excluded under the BBN analysis since for all
parameter regions their early-universe effect is huge.

In conclusion, $f(T,T_G)$ gravity, apart from having interesting cosmological 
implications both in inflationary and late-time phase, possesses particular 
sub-classes that can safely pass BBN bounds, nevertheless the torsional 
modification is constrained in narrow windows around the General Relativity 
values. This feature should be taken into account in future model 
building.

\section*{Acknowledgments}
This research is co-financed by Greece and the European Union (European Social 
Fund-ESF) through the Operational Programme “Human Resources Development, 
Education and Lifelong Learning” in the context of the project
“Strengthening Human Resources Research Potential via Doctorate Research” 
(MIS-5000432), implemented by the
State Scholarships Foundation (IKY). The work  of N.E.M is supported in part by 
the UK Science and Technology Facilities  research Council (STFC) under the 
research grant ST/T000759/1. 
S.B., N.E.M. and E.N.S. also acknowledge participation in the COST Association 
Action CA18108 ``{\it Quantum Gravity Phenomenology in the Multimessenger 
Approach (QG-MM)}''.

\end{document}